\newcommand{\ale}{\ \raisebox{-.3ex}{$\stackrel{<}{\scriptstyle \sim}$}\ }

\documentstyle{mn}

\input psfig.sty

\title[Self-gravitating protoplanetary discs]
	{Astrometric signatures of self-gravitating \\ protoplanetary discs}

\author[Rice et al.]{W.K.M. Rice$^1$, P.J. Armitage$^{2,3}$, M.R. Bate$^4$ and I.A. Bonnell$^1$ \\
	$^1$School of Physics and Astronomy, University 
	of St Andrews, North Haugh, St Andrews KY16 9SS \\
	$^2$JILA, Campus Box 440, University of Colorado, Boulder CO 80309-0440, USA \\
	$^3$Department of Astrophysical and Planetary Sciences, University of Colorado, Boulder CO 80309-0391, USA \\ 
	$^4$School of Physics, University of Exeter, Stocker Road, Exeter EX4 4QL}		
\begin{document}

\maketitle

\begin{abstract}
We use high resolution numerical simulations to study whether gravitational instabilities 
within circumstellar discs can produce astrometrically detectable motion of the
central star. For discs with masses of $M_{\rm disc} = 0.1 \ M_*$, 
which are permanantly stable against fragmentation,  
we find that the magnitude of the astrometric signal depends upon the 
efficiency of disc cooling. Short cooling times produce prominent 
filamentary spiral structures in the disc, and lead to stellar motions 
that are potentially observable with future
high precision astrometric experiments. For a disc that is marginally 
unstable within radii of $\sim 10 \ {\rm au}$, we estimate astrometric displacements 
of $10-10^2 \ \mu {\rm arcsec}$ on decade timescales for a star at a distance of 100~pc.
The predicted displacement is suppressed by a factor of several in more stable discs 
in which the cooling time exceeds the local dynamical time by an order of magnitude. 
We find that the largest contribution comes from material in the outer regions of the disc
and hence, in the most pessimistic scenario, the stellar motions caused by the disc 
could confuse astrometric searches for low mass planets orbiting at large radii. They are, 
however, unlikely to present any complications in searches for embedded planets orbiting at 
small radii, relative to the disc size, or Jupiter mass planets or greater orbiting at 
large radii.  
\end{abstract}


\section{Introduction}
High precision astrometry is a powerful tool to search for companions to nearby 
stars.  It also has the potential to discover significant numbers of extrasolar planetary 
systems.  In this paper we discuss the potential of astrometry as a probe of 
self-gravitating discs around pre-main sequence stars.

A gaseous disc with sound speed $c_s$, surface density $\Sigma$, and 
epicyclic frequency $\kappa$ is described as self-gravitating if the 
Toomre (1964) $Q$ parameter, 
\begin{equation} 
 Q = { {c_s \kappa} \over {\pi G \Sigma} },
\end{equation} 
is of order unity. In discs where self-gravity is important, 
the outcome can either be fragmentation into one or more bound objects, or 
a quasi-steady state in which gravitational instabilities 
lead to the outward transport of angular momentum. 
Local simulations suggest that the boundary between these 
possibilities is set by the ratio of the local dynamical time-scale $\Omega^{-1}$ 
to the time-scale on which the disc radiates thermal energy. 
Fragmentation occurs whenever the cooling time 
$t_{\rm cool} \ale 3 \Omega^{-1}$, while longer cooling times 
lead to stable angular momentum transport (Gammie 2001).

Circumstantial evidence suggests that self-gravity could 
play a role in protoplanetary discs as late as the optically visible 
Classical T Tauri phase, which lasts 
for several Myr (Haisch, Lada \& Lada 2001). Evidence that relatively 
old disc may be self-gravitating comes, first, from models of FU Orionis 
outbursts, which require a low efficiency of disc angular momentum transport 
to reproduce the observed $\sim 10^2 \ {\rm yr}$ time-scales. 
If the viscosity is parameterized using the Shakura-Sunyaev (1973) 
prescription, $\nu = \alpha c_s h$, where $h$ is the vertical scale height, 
FU Orionis models suggest 
a quiescent $\alpha \sim 10^{-4}$ (Bell \& Lin 1994; Bell et al. 1995). 
For a given accretion rate, small values of $\alpha$ imply high surface 
densities, so that the disc would be self-gravitating at $r \sim 1 \ {\rm au}$.
Second, theory suggests that angular momentum transport ought to 
be suppressed in cool regions of the disc where the gas is poorly 
coupled to magnetic fields (Matsumoto \& Tajima 1995; Gammie 1996; 
Fleming, Stone \& Hawley 2000; Sano et al. 2000; Reyes-Ruiz 2001; 
Sano \& Stone 2002). Again, this suggests that self-gravity may set 
in at radii of a few au as the first significant non-magnetic source 
of angular momentum transport (Armitage, Livio \& Pringle 2001). 
Ascertaining when self-gravity is at work within the disc requires 
either the observation of spiral patterns using extremely high resolution 
imaging, or detection of the astrometric motion of the stellar 
photocentre induced by the self-gravitating disc. It has been shown (Adams et al. 1989) 
that self-gravitating perturbations with $m=1$, can force the central 
star to move from the centre of mass.

In this paper, we use numerical simulations to quantify the 
magnitude of the astrometric displacement. This has previously been 
studied by Boss (1998), who simulated a disc with a mass of 
$\approx 0.2 M_*$ and found a large motion of the star, of the 
order of 0.1~au. This corresponds to milliarcsecond displacements at 
the distance of the nearest star-forming regions, which would be 
easily detectable by any of the forthcoming high precision 
astrometry experiments. 
The disc simulated by Boss (1998), however, was highly unstable, 
and subsequently fragmented with the formation of substellar 
objects. Although promising for giant planet formation (Armitage \& 
Hansen 1999; Boss 2000), prompt fragmentation implies that extremely 
fortuitous timing would be needed for the astrometric detection of 
self-gravitating discs. We concentrate instead on marginally 
unstable discs, which are not vulnerable to fragmentation and 
could potentially exist around many  
Classical T Tauri stars. 

\section{Numerical simulations}

\subsection{Smooth particle hydrodynamics code}
The three-dimensional simulations presented here were performed using smooth 
particle hydrodynamics (SPH), a Lagrangian hydrodynamics code 
(e.g., Benz 1990; Monaghan 1992).  Our implementation allows for the inclusion of point masses and
for point mass creation (Bate et al. 1995). In this simulation the central 
star is modelled as a point mass onto which gas particles can accrete if they
approach to within the sink radius. Although point mass creation is allowed,
the discs considered here are stable against fragmentation and the density never
reaches values high enough for the creation of a point mass within the disc
itself. A great saving in computational time is made by using individual,
particle time-steps (Bate et al. 1995; Navarro \& White 1993).  The time-steps
for each particle are limited by the Courant condition and a force condition
(Monaghan 1992).  Both point masses and gas use a tree to determine neighbours 
and to calculate gravitational forces (Benz et al. 1990). 

An advantage of using SPH for this calculation is that it is not necessary to impose
any outer boundary conditions, and the SPH particles are free to move to radii 
greater than the initial disc radius. The outer edge of the disc is therefore free to
distort and modes with $m=1$ will not be affected or artificially driven by the outer 
boundary conditions (Heemskerk et al., 1992).  

\subsection{Initial conditions}
We consider a system comprising a central star, modelled as a point mass with mass $M_*$, 
surrounded by a gaseous
circumstellar disc with a mass of $0.1 M_*$.  The disc temperature is taken to
have an initial radial profile of $T \propto r^{-0.5}$ (e.g. Yorke \& Bodenheimer 1999) 
and the Toomre $Q$ parameter is
assumed to be initially constant with a value of $2$. A stable accretion disc where 
self-gravity leads to the
steady outward transportation of angular momentum should have a near constant Q
throughout.  A constant $Q$ together with equation (1) then
gives a surface density profile of $\Sigma \propto r^{-7/4}$, and hydrostatic
equilibrium in the vertical direction gives a central density profile of 
$\rho \propto r^{-3}$.  

The disc is modelled using 250,000 SPH
particles, which are initially distributed according to the specified 
density profile between inner and outer radii of $r_{\rm in}$ and 
$r_{\rm out}$ respectively. The actual particle positions are chosen
randomly, subject to the constraint that the centre of mass (and momentum) 
of the disc is initially coincident with that of the star. To achieve 
this, we simply add particles in point symmetric pairs about the central 
point mass. 

The calculations performed here are essentially scale free. In code units, 
we take $M_* = 1$, $r_{\rm in} = 1$ and $r_{\rm out} =25$. To simplify the 
discusion we will generally assume a physical mass scale of $1 M_{\odot}$ 
and a length scale of $1$ au.
The circumstellar disc therefore has a mass of $0.1 M_{\odot}$, extends from 
1~au to 25~au, and surrounds a
star of mass $1 M_{\odot}$.

\subsection{Cooling}
Previous work has shown that the outcome of gravitationally unstable 
discs depends critically on the treatment of the energy equation 
(Pickett et al. 2000; Gammie 2001). Our global simulations adopt 
the same approach as was used for local models by Gammie (2001). 
We use an adiabatic equation of state, with adiabatic index $\gamma = 5/3$, and 
allow the disc gas to heat up due to both PdV work and viscous 
dissipation. In the absence of cooling, viscous dissipation 
would act to heat up the disc, increase the Toomre $Q$ parameter, 
and drive the disc towards stability.
Cooling is implemented by adding a simple cooling 
term to the energy equation. Specifically, for a particle with 
internal energy per unit mass $u_i$,
\begin{equation} 
 { {{\rm d} u_i} \over {{\rm d}t} } = - {u_i \over {t_{\rm cool}}}
\end{equation}  
where $t_{\rm cool}$ is a radially dependent parameter which we 
specify for each run.

Although simple, this approach to the energetics of the disc can 
be related (at least approximately) to the real physics of an 
accretion disc.
For an optically thick disc in equilibrium the cooling
time is the ratio of the thermal energy per unit area to the radiative losses
per unit area. It can be
shown (see e.g. Pringle 1981) that in such a viscous accretion disc, 
the cooling time is given by 
\begin{equation}
 t_{\rm cool} = \frac{4}{9\gamma}\frac{1}{\alpha \Omega}
\end{equation}
where $\gamma$ is the adiabatic index, $\Omega$ is the angular frequency, and
$\alpha$ is the Shakura \& Sunyaev (1973) viscosity parameter. Based on this, 
we assume radially dependent cooling times that scale as $\Omega^{-1}$,
\begin{equation}
 t_{\rm cool}=\beta \Omega^{-1},
\end{equation} 
with $\beta$ a constant. Gammie (2001) has
shown, using a local model, that cooling times of $t_{\rm cool} < 3
\Omega^{-1}$ lead to fragmention, while longer cooling times lead to a
quasi-stable state. Since we are interested in the possibility of
detecting, through the displacement of the central star, 
the presence of gravitational instabilities in quasi-stable circumstellar discs,
we will use $\beta$ values greater than 3. 

\section{Astrometric signal}
We consider a $0.1 M_{\odot}$  disc with an inner radius of 1~au 
and an outer radius of 25~au. The disc is initially gravitationally stable
($Q\sim 2$ at all radii) and undergoes cooling with a radially dependent 
cooling time given by $t_{\rm cool}=\beta \Omega^{-1}$. We consider cooling times of
$5 \Omega^{-1}$ and $10 \Omega^{-1}$ which, according to Gammie (2001), should
lead to quasi-stable discs that do not fragment. The $t_{cool}=10 \Omega^{-1}$ and 
$t_{cool}=5 \Omega^{-1}$ simulations were run for 2.6 and 3.7 orbits of particles at 
the outer disc edge at $r_{\rm out} = 25$. 
These run times are short (326 years and 467 years respectively for our assumed scaling) 
compared to the evolutionary time of the disc (i.e the viscous time, $r^2 / \nu$) but 
are sufficient to reach a quasi-steady state in which heating and cooling are 
locally in balance. 

For a cooling time of $10 \Omega^{-1}$ (Figure 1) there is no noticeable structure in the disc. For
a cooling time of $5 \Omega^{-1}$ (Figure 2) there is significant structure in the disc but, as
in Gammie (2001), there has been no fragmentation. The filamentary spiral structure which 
we observe is consistent (e.g. Nelson et al. 1998) with previous simulations of discs 
with masses significantly less than that of the
central star. Although it is likely that the $m=1$ mode is the dominant mode 
driving the stellar
motion (Adams et al. 1989), it is not an obviously dominant mode in the
disc.

\begin{figure}
\centerline{\psfig{figure=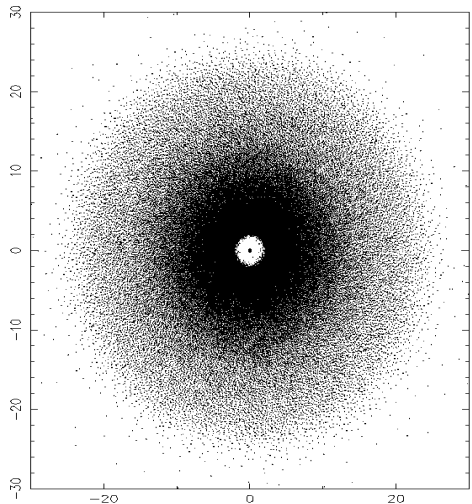,width=3.0truein,height=3.0truein}}
\caption{\label{tc10disk} Face on view of disc with $t_{cool}=10 \Omega^{-1}$ after a time of
326 years. There is no noticeable structure in the disc.}
\end{figure}

\begin{figure}
\centerline{\psfig{figure=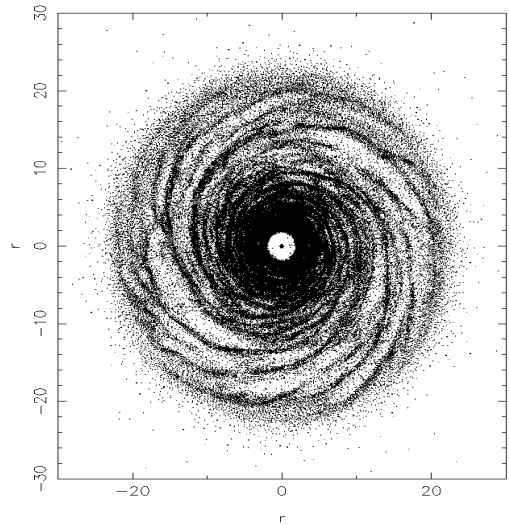,width=3.0truein,height=3.0truein}}
\caption{\label{tc5disk} Face on view of disc with $t_{cool} = 5 \Omega^{-1}$ after a time of 467
years. There is significant structure in the disc, but the maximum density at this time
is not significantly different from the maximum density at the beginning of the simulation, and
no fragmentation has occurred.}
\end{figure}

Figures 3 and 4 show the surface densities of the $t_{cool}= 10 \Omega^{-1}$ and $t_{cool}= 5
\Omega^{-1}$ simulations at the beginning and end of each run. Because we have not 
attempted to model the inner boundary condition of the disc in any detail, there is 
rapid accretion and a drop in surface density close to the inner boundary. As discussed 
below, we find that the main contribution driving displacement of the star comes 
from material near the outer edge of the disc, so this numerical effect near the 
inner boundary is not a major concern. Apart from these particles with 
radii between 1 and 2 being accreted onto the central star, the surface density profile does not
change significantly during the course of the simulation. 

\begin{figure}
\centerline{\psfig{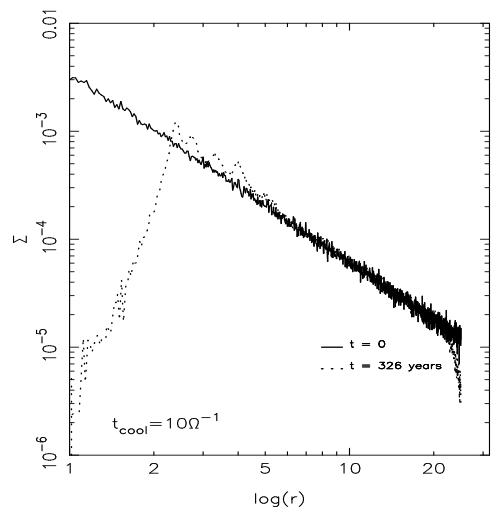}}
\caption{\label{tc10sigma} Surface density of disc with $t_{cool}=10  \Omega^{-1}$ at the beginning
($t=0$) and end ($t=326$ years) of the simulation.}
\end{figure}

\begin{figure}
\centerline{\psfig{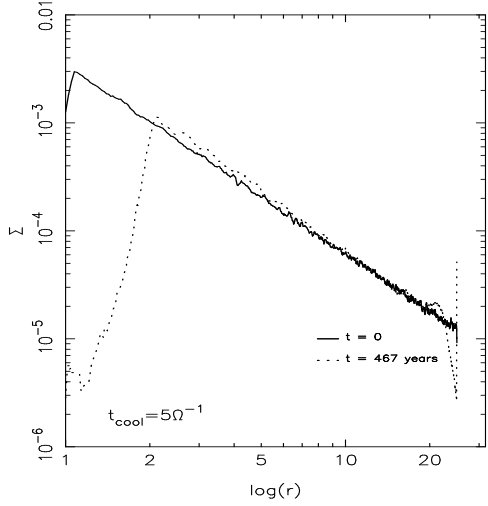}}
\caption{\label{tc5sigma} Surface density of disc with $t_{cool}=5  \Omega^{-1}$ at the beginning
($t=0$) and end ($t=467$ years) of the simulation.}
\end{figure}

Figures 5 and 6 shows the Toomre $Q$ parameter at the beginning and end of each simulation. 
For $t_{cool}=10 \Omega^{-1}$  the value of $Q$ at
the end of the simulation is smaller than the initial value of $2$ but still generally greater than
$1$. This disc is largely stable, although $Q$ is small enough
that some structure (not noticeable in Fig 1) may exist. For $t_{cool}=5
\Omega^{-1}$ the final $Q$ value is close to 1 for radii between 2 and 15. The disc is quasi-stable
with heating through gravitational instabilities balancing the cooling to give $Q \sim 1$. 

\begin{figure}
\centerline{\psfig{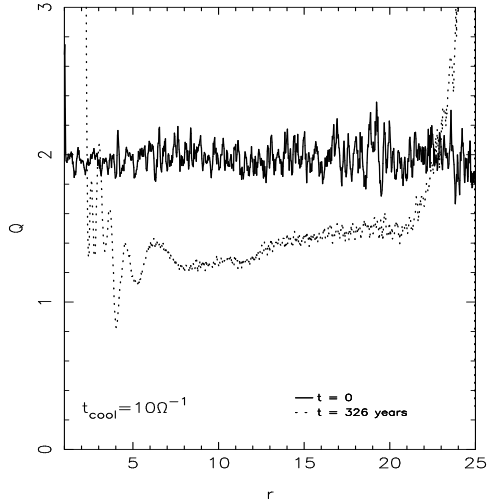}}
\caption{\label{tc10Q} Toomre $Q$ parameter for $t_{cool}=10  \Omega^{-1}$.  At the beginning
of the simulation ($t=0$) $Q$ has a constant value of 2. At the end ($t=326$ years) of the simulation
$Q$ largely lies between the intial value of $2$ and the critical value $1$.}
\end{figure}

\begin{figure}
\centerline{\psfig{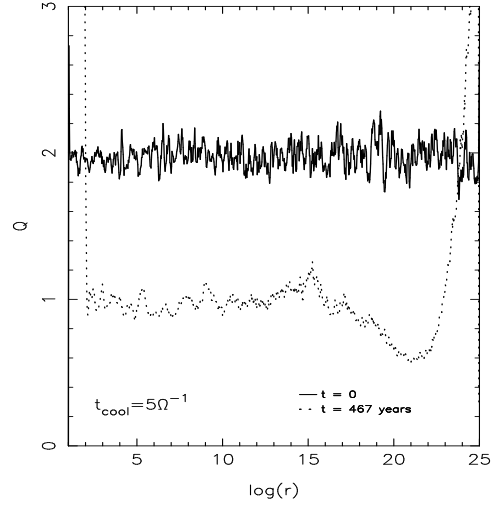}}
\caption{\label{tc5Q} Toomre $Q$ parameter for $t_{cool}=5  \Omega^{-1}$.  At the beginning
of the simulation ($t=0$) $Q$ has a constant value of 2. At the end ($t=467$ years) of the simulation
$Q$ has an almost constant value of $1$ between radii of 2 and 15.}
\end{figure}

Figures 7 and 8 show the displacement of the central star from the centre of mass of the 
star-disc system as a function of time, for the runs with $t_{\rm cool}=10 \Omega^{-1}$ and 
$t_{\rm cool}=5 \Omega^{-1}$ respectively. The displacement grows approximately exponentially 
at early times, before saturating (Laughlin et al. 1997)
and reaching a plateau at a displacement which 
depends upon the assumed cooling time. For the more unstable disc, with $t_{\rm cool}=5 \Omega^{-1}$, 
the saturation level of the stellar displacement is $\sim 5 \times 10^{-3}$ au, 
while for the more stable disc with $t_{\rm cool}=10 \Omega^{-1}$ the plateau occurs 
at a lower level of $\sim 10^{-3}$ au. For a source at 100 pc, the above displacements will
produce angular displacements of 0.1 milliarcsec (mas) and 0.02 mas respectively. 
In both cases the star is executing 
an approximately circular orbit around the centre of mass, with periods of 
approximately 50~yr (for $t_{\rm cool}=5 \Omega^{-1}$) and 35~yr (for $t_{\rm cool}=10 \Omega^{-1}$).

Instead of the disk extending from 1 - 25 au, the scale free nature of the simulation
means that we could equally well assume, without changing the mass scale, 
that it extends from 4 - 100 au, closer to a 
more commonly observed
disc size (Padgett et al. 1999). The displacement of the
central star from the center of mass and the angular displacement would both be 
4 times greater, and the period of the orbit would be 8 times 
greater than that of a 25 au disk. A planet producing the same displacement with 
the same period as that calculated here, would have a mass several times less
than that of Jupiter. A Jupiter mass planet producing the same displacement 
would be significantly closer to the central star and hence
would have a correspondingly smaller period. It therefore seems unlikely (unless considering low
mass planets at large radii) that there could be much confusion between astrometric signals 
caused by planets and those due to disc instabilities. 

In addition to a roughly circular motion, gravitational instability in the 
disc also generates lower amplitude, short-timescale motions. This is 
illustrated in Figure~9, which shows a projection of the position of the central star 
onto the orbital plane for the last 57 years of the simulation with $t_{\rm cool}=5 \Omega^{-1}$. 
The center of mass is located at $x=0, y=0$. As well as
the large orbit taking $\sim 50$ years there are a number of smaller orbital structures
due to the presence of higher order modes in the disc.

For our surface density profile, most of the mass in the disc lies at relatively 
large radius (the enclosed mass scales as $r^{1/4}$). This behaviour is even 
stronger for the flatter surface density profiles (typically $r^{-1}$ or $r^{-3/2}$) 
often considered as appropriate for protoplanetary discs. As expected given 
this mass distribution,  
the timescale of the dominant circular motion of the star is of the same order 
as the dynamical timescale in the {\em outer} disc. Assuming this to be true 
more generally, we can obtain estimates for the angular displacement $\Delta \theta$ 
and characteristic timescale $\tau$ as a function of the outer radius, $r_{\rm grav}$, 
within which a disc is self-gravitating. For a star at distance $d$ we find,
\begin{eqnarray}
 \Delta \theta & \approx &  100 \left( {r_{\rm grav} \over {25 \ {\rm au}} } \right) 
 \left( { d \over {100 \ {\rm pc} } } \right)^{-1} \ \mu {\rm arcsec} \\ 
 \tau & \approx & 50 \left( {r_{\rm grav} \over {25 \ {\rm au}} } \right)^{3/2} \ {\rm yr}, 
\end{eqnarray}
where we have used the numbers from the $t_{\rm cool}=5 \Omega^{-1}$ run for the 
estimate. We expect the exact numbers to depend upon the actual surface 
density distribution in the disc, and on the mass, so these figures should be regarded as 
order of magnitude estimates for discs with masses around a tenth of the 
mass of the star.
 
We can also compare the displacement caused by a gravitionally unstable disc 
with that generated by an orbiting planet within the disc. For a planet 
whose mass ratio to the central star is $q = M_p / M_*$, in a circular 
orbit at radius $a$, the displacement is,
\begin{equation} 
 \Delta \theta = 50 \left( {q \over 10^{-3}} \right)
 \left( {a \over {5 \ {\rm au}} }\right) 
 \left( { d \over {100 \ {\rm pc} } } \right)^{-1} \ \mu {\rm arcsec},
\end{equation} 
i.e. of similar magnitude to the result derived above. The {\em period} 
of the stellar oscillation caused by the planet, however, is substantially 
shorter than that generated by the disc. We conclude, therefore, that even 
in the worst case scenario where (i) the disc is gravitationally unstable, 
and (ii) the cooling time is within a factor of two of the fragmentation 
boundary, stellar displacements due to the disc are {\em not} likely 
to be confused with the signal from a Jupiter mass planet. At 
substantially lower masses, of course, confusion would be possible. 
In this case, detection of the substructure in the orbit shown 
in Figure~9 would be necessary to unambiguously determine whether 
an observed signal was of disc or planetary origin. The disc mass of $0.1$ M$_*$
is also quite large, with most T Tauri stars having disc masses
smaller than that used here (Beckwith et al. 1990). 
Not only will lower mass discs be less likely to be 
gravitationally unstable (see Equation (1)), 
the amplitude of the displacement of the central star is likely to be 
smaller than that obtained here. This will further reduce the 
possibility of confusing stellar
displacements due to the disc with that due to an orbiting planet.

\section{Conclusion}
We have used numerical simulations of discs around Young Stellar Objects 
to quantify the astrometric displacement of the star caused by a 
self-gravitating disc. By modelling the energy balance of the disc 
using a cooling time formalism (Gammie 2001), we have shown that 
the magnitude of the displacement depends upon how stable the 
disc is against fragmentation into bound substellar objects. For a 
cooling time of $5 \Omega^{-1}$ (within a factor of two of the 
fragmentation boundary at $\approx 3 \Omega^{-1}$), a disc 
mass of $0.1 \ M_*$, and a self-gravitating disc radius of 
the order of 10~au, we obtain relatively large displacements 
in the $10-10^2 \ \mu {\rm arcsec}$ range (for a star at a
distance of 100~pc). This magnitude of astrometric signal is 
potentially observable with any of several upcoming high 
precision astrometric instruments, although the presence of a circumstellar
disc is likely to complicate such observations, especially if time dependent
perturbations are present within the disc. Although the gravitationally driven
instabilities modelled in this work are intrinsically time dependent, their growth
saturates (Laughlin et al. 1997) and their filamentary nature (Nelson et al. 1998) 
makes their structure 
approximately azimuthally symmetric. Their presence is therefore unlikely to further
complicate observations. Discs with  
cooling times longer than $5 \Omega^{-1}$, which are correspondingly more stable, 
have significantly smaller stellar motions. 
A detailed model for the angular momentum transport processes in the disc, 
together with an assessment of the heating and cooling processes at work, 
would be required to determine whether protoplanetary discs fall into 
the parameter regime that produces the largest displacements.

Our estimate for the angular displacement of the star caused by a self-gravitating 
disc is an order of magnitude smaller than that found by Boss (1998). Although 
there are numerous differences in the initial conditions and numerical 
techniques used, we believe that the smaller displacement that we obtain primarily 
reflects our choice of a more stable disc model. By design, our disc is 
permanently stable against fragmentation into substellar objects, whereas 
the disc simulated by Boss (1998) broke up towards the end of the simulation. 
Taken together, our results, plus those of Boss (1998), suggest that a 
relatively narrow window of self-gravitating disc conditions could lead to 
small but long-lived astrometric wobbles ($10-10^2 \ \mu {\rm arcsec}$ 
at 100~pc). More dramatic displacements -- perhaps of the order of a mas at 
the same distance -- are also possible, but only as a precursor to 
fragmentation of the disc into substellar objects. Our simulations suggest that
the disc induced wobble could mimic that of a planet a few times smaller than Jupiter 
and orbiting at a large radius, relative to the disc size. 
Most T Tauri stars have disc masses 
smaller than that used in this simulation and would consequently induce a correspondingly
smaller motion of the central star, reducing  the likelihood of confusion between
stellar motion due to disc instabilities and that due to an orbiting planet.   
The disc 
induced stellar motions are therefore not likely to be a serious obstacle 
to the astrometric detection of planets orbiting at small radii (relative to the disc
size). For planets orbiting at large radii, the planet mass which could be mimiced by
a disc instability depends on the disc mass (relative to the central star) but
is likely to be less than a Jupiter mass for most T Tauri disc systems.

\begin{figure}
\centerline{\psfig{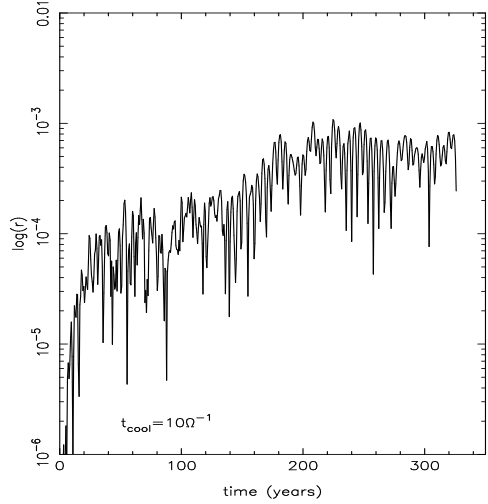}}
\caption{\label{tc10rt} Time evolution of the distance of the central star from the center of mass
for $t_{cool}=10 \Omega^{-1}$. At the end of the simulation the system seems to have settled into a
steady state in which the central star is orbiting the center of mass with a orbital radius of $\sim
10^{-3}$ radial units.}
\end{figure}

\begin{figure}
\centerline{\psfig{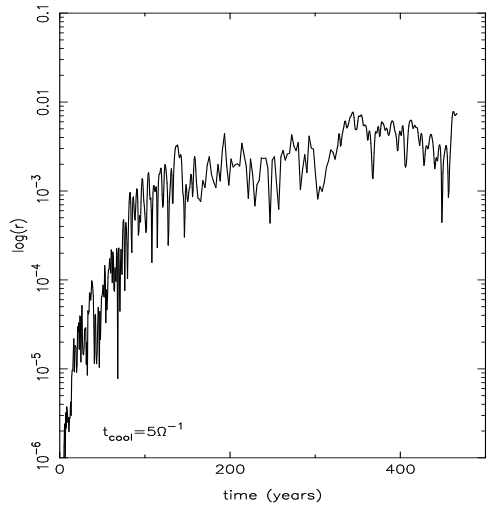}}
\caption{\label{tc5rt} Time evolution of the distance of the central star from the center of mass
for $t_{cool}=5 \Omega^{-1}$. At the end of the simulation the system seems to have settled into a
steady state in which the central star is orbiting the center of mass with a orbital radius of $\sim
5 \times 10^{-3}$ radial units.}
\end{figure}

\begin{figure}
\centerline{\psfig{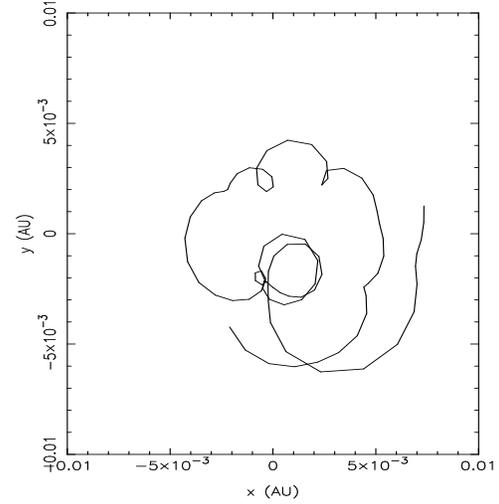}}
\caption{\label{tc5xy} Projection of the position of the central star onto the orbital plane.  The
center of mass is at $x=0, y=0$. The orbit is approximately circular and has substructure due to
perturbations from higher order modes.}
\end{figure}

\section*{Acknowlegments}

The simulations reported in this paper made use of the 
UK Astrophysical Fluids Facility (UKAFF).

\end{document}